\documentclass[11pt]{article}

\begin{document}

\input{epsf}

\def\beq{\begin{equation}}
\def\eeq{\end{equation}}
\def\bea{\begin{eqnarray}}
\def\eea{\end{eqnarray}}
\def\beas{\begin{eqnarray*}}
\def\eeas{\end{eqnarray*}}
\def\ov{\overline}
\def\ot{\otimes}

\newcommand{\hf}{\mbox{$\frac{1}{2}$}}
\def\sig{\sigma}
\def\De{\Delta}
\def\af{\alpha}
\def\be{\beta}
\def\la{\lambda}
\def\ga{\gamma}
\def\ep{\epsilon}
\def\vep{\varepsilon}
\def\half{\frac{1}{2}}
\def\third{\frac{1}{3}}
\def\fth{\frac{1}{4}}
\def\sth{\frac{1}{6}}
\def\tth{\frac{1}{24}}
\def\tde{\frac{3}{2}}

\def\zb{{\bar z}} 
\def\psib{{\bar \psi}} 
\def\etab{{\bar \eta }}
\def\gab{{\bar \ga}}
\def\vev#1{\langle #1 \rangle}
\def\inv#1{{1 \over #1}}

\def\CA{{\cal A}}       \def\CB{{\cal B}}       \def\CC{{\cal C}}
\def\CD{{\cal D}}       \def\CE{{\cal E}}       \def\CF{{\cal F}}
\def\CG{{\cal G}}       \def\CH{{\cal H}}       \def\CI{{\cal J}}
\def\CJ{{\cal J}}       \def\CK{{\cal K}}       \def\CL{{\cal L}}
\def\CM{{\cal M}}       \def\CN{{\cal N}}       \def\CO{{\cal O}}
\def\CP{{\cal P}}       \def\CQ{{\cal Q}}       \def\CR{{\cal R}}
\def\CS{{\cal S}}       \def\CT{{\cal T}}       \def\CU{{\cal U}}
\def\CV{{\cal V}}       \def\CW{{\cal W}}       \def\CX{{\cal X}}
\def\CY{{\cal Y}}       \def\CZ{{\cal Z}}

\newcommand{\np}{Nucl. Phys.}
\newcommand{\pl}{Phys. Lett.}
\newcommand{\prl}{Phys. Rev. Lett.}
\newcommand{\cmp}{Commun. Math. Phys.}
\newcommand{\jmp}{J. Math. Phys.}
\newcommand{\jpamg}{J. Phys. {\bf A}: Math. Gen.}
\newcommand{\lmp}{Lett. Math. Phys.}
\newcommand{\ptp}{Prog. Theor. Phys.}

\newif\ifbbB\bbBfalse                
\bbBtrue                             

\ifbbB   
 \message{If you do not have msbm (blackboard bold) fonts,}
 \message{change the option at the top of the text file.}
 \font\blackboard=msbm10 
 \font\blackboards=msbm7 \font\blackboardss=msbm5
 \newfam\black \textfont\black=\blackboard
 \scriptfont\black=\blackboards \scriptscriptfont\black=\blackboardss
 \def\Bbb#1{{\fam\black\relax#1}}
\else
 \def\Bbb{\bf}
\fi

\def\id{{1\! \! 1 }}
\def\bo{{\Bbb 1}}
\def\bI{{\Bbb I}}
\def\bC{{\Bbb C}} 
\def\bZ{{\Bbb Z}}
\def\bR{{\Bbb R}}
\def\CN{{\cal N}}

\title{Series expansions for lattice Green functions}
\author{{\bf Z. Maassarani}\thanks{Work supported  by DOE grant no.
DE-FG02-97ER41027 and NSF grant no. DMR-9802813.} \\
\\
{\small Department of Physics}\\
{\small University of Virginia}\\
{\small 382 McCormick Road}\\
{\small P.O. Box 400714}\\
{\small Charlottesville, VA,  
22904-4714 USA}\thanks{Email address: zm4v@virginia.edu} \\}
\date{}
\maketitle

\begin{abstract}
Lattice Green functions appear in  lattice gauge theories,
in  lattice models of statistical physics and in random walks. 
Here, space coordinates are treated as parameters and series expansions in 
the mass are obtained. The singular points in arbitrary dimensions 
are found. For odd dimensions these are branch points
with half odd-integer exponents, 
while for even dimensions they are of the logarithmic type.  
The  differential equations for one, two and three
dimensions are derived, and the general form for 
arbitrary dimensions is indicated. Explicit series expressions are 
found in one and two dimensions. These series are
hypergeometric functions. In three and higher dimensions the series
are more complicated. 
Finally an algorithmic  method by  Vohwinkel, L\"uscher 
and Weisz is shown to generalize to arbitrary anisotropies and mass. 
\end{abstract}
\vspace*{2.5cm}
\noindent
\hspace{1cm} March 2000\hfill\\

\thispagestyle{empty}

\newpage

\setcounter{page}{1}

\section{Introduction}

Green functions, also  known as two-point functions  or propagators,
are ubiquitous in physics. They appear in 
classical field theories, quantum mechanics and quantum field theories 
to name but a few areas. When a discrete space-time approach is used 
as a regulator, the  continuum Green functions become  lattice functions.
In the problems of  condensed matter physics
both the continuum and lattice approach are natural.  
Random walks on a lattice provide another field of application for Green
functions, where they are generating functions for
the return probability of a random walker on a lattice \cite{hughes}. 
Given an isotropic random walk occurring in steps of one unit on a hypercubic
lattice, one considers the probability of visiting a given site, or of
returning to the starting site, after $n$ steps. 
This problem was first considered by P\'olya \cite{polya} who showed
that this probability tends to one only in one and two dimensions, 
for $n$ tending to infinity.
This is the recurrence/transience transition which is  
an important result in the study of random walks.  

The free propagator appears naturally in the perturbative 
expansion of field theories.
A lattice  provides a regulator for the 
infinities in such expansions, 
and the  four-dimensional discrete Green functions appear 
naturally \cite{creutz,rothe,montvay}.
The three-dimensional function appears in the   study of effective  
three-dimensional theories \cite{rumope}. 
A lattice regularization has also
been used in \cite{adhi} to study non-relativistic quantum scattering
in two and three space dimensions. 
For dimensions greater than one,
the continuum free Green function is singular at the origin, 
while the  lattice version  is finite. This provides a regularization
of the UV divergences. 
In yet another context,  the propagator was used to find the resistance 
of a network of resistors \cite{cser}. (See also \cite{henkar}
in relation to conformal invariance.)

Although the integral representation of the lattice Green function
for the hypercubic lattice is well-known, the analytic calculation of this
integral remains a challenge. A closed form appears to exist only 
in one dimension and for a special case in two dimensions. 
In higher dimensions there are several partial results \cite{hughes,glasser}. 
A rather recent  numerical approach was used in \cite{luwe1}
and developed in \cite{luwe2} for the massless free propagator. 
It was applied to the two-dimensional case in \cite{shin}. 
Similar position space methods were considered earlier in \cite{frimar},
and later in \cite{palasex}. 

In this paper I consider the free propagator as a function of the 
mass squared which is allowed to vary in the whole complex plane. 
The space coordinates are treated  as parameters. 
Series expansions are first obtained for large values of the mass. In 
one and two dimensions these series are hypergeometric while in higher
dimensions they are not. This large mass 
expansion follows from the multidimensional 
integral representation of the lattice Green function. However this 
representation does not allow one to find the series expansion 
around other values of the mass. To this end recurrence relations
and corresponding differential equations
in the mass squared are derived. This approach gives  the singularities 
of the Green function and allows one to  expand around any point. 
Non-trivial monodromies are found around the singular points. 

The paper is organized as follow. Section 2 introduces the notation, and   
the well-known integral representation  for the minimal anisotropic 
lattice Green function for the simple cubic lattice in 
$d$ dimensions. Some general  properties are also discussed.  
The  one-dimensional integral representation in terms of  
Bessel functions  is given.
A simple derivation for the location of the singularity in the complex mass
plane is found. A general expansion of the lattice Green function 
in the inverse of the mass is obtained, and a recurrence relation
between dimensions for the coefficients of this expansion is noted. 
In section 3 the one-dimensional case is studied in detail
as a simple illustration of the methods used. This also serves as a guide
to the features common to all dimensions. The same methods
are  applied to the two dimensional Green function and two novel
expansions are found (section 4). The recurrence relation and 
the differential equation are then obtained for 
the three-dimensional case (section 5). 
Section 6 concludes with general remarks for four and higher dimensions.
Appendix A contains a set of formul\ae \/ used in the text.  
In appendix B,  the method of \cite{luwe2} is extended  
to arbitrary mass and anisotropies.

\section{General results}\label{genres}

Consider the hypercubic lattice in $d$ dimensions, with unit vectors
$\hat{e}_j$, $j=1,\cdots, d$. 
The  discrete anisotropic Green equation is 
\beq
H f(\vec{x}) \equiv \sum_{j=1}^d \alpha_j
\left[ f(\vec{x}+\hat{e}_j)+f(\vec{x}-\hat{e}_j) \right] =
2\be\,  f(\vec{x}) + \delta_{\vec{x},\vec{0}} \label{greq}
\eeq
The integers $x_j$ label the lattice sites, and the anisotropies 
$\alpha_j$ are arbitrary but non-vanishing complex numbers. 
When the $\alpha_j$'s are omitted, they should be assumed to 
be all equal to 1. 
The action of the discrete anisotropic $d$-dimensional Laplace 
operator $\Delta$ on $f$ is given by $(H -2 
\sum_{j=1}^d \alpha_j )\, f(\vec{x})$.   
Here the contribution of the latter term has been absorbed in $\be$ 
which is, depending on the context, the mass squared, the energy eigenvalue
or a formal expansion parameter for a generating function.  
This definition of  $\be$ gives a natural  parity symmetry (\ref{sym1}), 
and renders the sets of singularities symmetric with respect to the origin.
The free massless scalar propagator of lattice gauge theories corresponds
to the isotropic case, $\alpha_j=1$ for all $j$, and  $\beta=d$. 
A completely  isotropic solution of the isotropic equation satisfies
\beq
2d\, f(\hat{e}_j) = 2\be\, f(\vec{0}) + 1\; , \quad \quad j=1,\cdots,d
\eeq
Adding to a solution of (\ref{greq}) 
any solution of the homogeneous equation,
\beq
\sum_{j=1}^d \alpha_j
\left[ f(\vec{x}+\hat{e}_j)+f(\vec{x}-\hat{e}_j) \right] =
2\be\,  f(\vec{x})  \; ,\label{greh}
\eeq
yields another Green function. 
A large class of such solutions can be written as
\beq
h(\vec{x}|\vec{\alpha},\be)= 
\int_{-\pi}^{+\pi}\cdots \int_{-\pi}^{+\pi}
\frac{d^d\vec{q}}{(2\pi)^d} \exp (i \vec{q}\cdot\vec{x})\,\,\tilde{h}(\vec{q})
\,\,\delta(\sum_{j=1}^d \alpha_j\cos q_j -\be)
\eeq
where the function $\tilde h$ is arbitrary but well-behaved.

One can define a {\it decoupling point} at $\be=0$, 
for which equation (\ref{greq})
becomes an equation for two sub-lattices. A lattice where $\sum_{j=1}^d x_j$ 
is even and one where $\sum_{j=1}^d x_j$ is odd. 

The minimal solution of (\ref{greq}) is given by
\beq
G^{(d)}_{\pm}(\vec{x}|\vec{\alpha},\be)
=\int_{-\pi}^{+\pi}\cdots \int_{-\pi}^{+\pi}
\frac{d^d\vec{q}}{(2\pi)^d} \,\frac{\exp (i \vec{q}\cdot\vec{x})}{2
\sum_{j=1}^d \alpha_j\cos q_j - 2 \be \pm i \epsilon} \label{gref}
\eeq
where $\epsilon \rightarrow 0^+$.
The $\epsilon$ prescription removes integration ambiguities at the poles. 

When only $d'$ anisotropy parameters are equal to each other, 
the solution (\ref{gref}) is invariant under the
$2^d \cdot d'!$ parity transformations and permutations of the $x_j$'s. 
For $d'=d$, the symmetry is that of the $d$-dimensional hypercubic group, 
and the functions $G^{(d)}_{\pm}$ are completely symmetric 
in the absolute values of their  arguments $x_j$.
There is also a parity symmetry relating $\be$ to $-\be$,
\beq
G^{(d)}_{\pm}(\vec{x}|\vec{\alpha},-\be)= 
e^{i\pi\left(1+X\right)}\,
G^{(d)}_{\mp}(\vec{x}|\vec{\alpha},\be)\label{sym1}
\eeq
where $X\equiv\sum_{j=1}^d |x_j|$,
and a complex conjugation symmetry
\beq
\left( G^{(d)}_{\pm}(\vec{x}|\vec{\alpha},\be)\right)^*= 
G^{(d)}_{\mp}(\vec{x}|\vec{\alpha}^*,\be^*) \label{sym2}
\eeq
The latter symmetry shows that the two Green functions are complex conjugates
of each other. 

The  exponentiation formula 
\beq
\frac{n!}{A^{n+1}}=\int_0^\infty dt\, t^n \exp(-t A)   \;\;, \quad\quad
{\rm Re}(A) > 0 \; , \quad n=0,1,2,\cdots \label{expo}
\eeq
and the integral representations (\ref{intrep}) for the Bessel functions
yield a one-dimensional integral representation for (\ref{gref}):
\beq
G^{(d)}_{\pm}(\vec{x}|\vec{\alpha},
\be)=-\frac{(\pm i)^{1+X}}{2}
\int_0^{+\infty} dt \exp\left(-\frac{t\epsilon}{2} \mp i 
t\be\right) J_{|x_1|}(\alpha_1 t) \cdots J_{|x_d|}(\alpha_d t) \label{greb}
\eeq
In this expression $\epsilon$ can be set to zero. For $\alpha_j\in \bR$,
the integrals converge in the domains $\mp{\rm Im}\be \geq 0$, without  
a finite number of real points. 
Note that (\ref{greb}) satisfies (\ref{greq}) by virtue of properties
(\ref{prop1}--\ref{prop4}). 

To find the singularities  of $G^{(d)}_{\pm}$ in $\be$, consider 
first the initial integral representation (\ref{gref}). 
For $|\be|>\sum_j |\alpha_j|$, the integrand is a continuous function, without 
singularities, integrated over a compact domain. Thus the Green function
has no singularities for these values of $\be$. This includes the point
at infinity. 
Now consider, for  simplicity, 
the case without anisotropies ($\alpha_j=1$). Equation (\ref{gref}) implies
that the possible singularities are real.  
When $d=1$,  the integral $(\ref{greb})$ converges 
provided the oscillating cosine of the asymptotic expansion 
(\ref{assbes}) is not ``canceled''
by  $\exp(\mp it\be)$. For $\be = \pm 1$, and only for these values, 
the integral has a diverging contribution of the 
form $\int^\infty dt/\sqrt{t}$, which results in the 
branch points  $\be=\pm 1$. 
This is confirmed by the explicit expressions given in section \ref{d1}.
The same reasoning holds  for $d=2$. The product of the
two cosines yields one divergent contribution, $\int dt/t$ for 
three values of $\beta$:  $0$ and $\pm 2$. This is  confirmed 
in section \ref{d2}. For $\be=0$, note that
this approach also predicts a lack of divergence 
for points  on the odd sub-lattice. One has $\be\ln\be$, which vanishes 
as $\be$ tends to 0. 
For $d\geq 3$, there are enough powers of $\sqrt{t}$ to give a converging
integral at all $\be$. This corresponds to the recurrence/transience 
transition in the context of random walks. Let $[n]$ denote the integer part
of $n$.  Taking $[(d-1)/2]$ $\be$-derivatives of (\ref{greb}),
and using the  asymptotic expansion for the 
Bessel function, yields  $d+1$ singularities: $\be_0=-d,-d+2,\cdots,d-2,d$. 
For $d\geq 1$ and odd, they  are of the branch point 
type:  $(\be-\be_0)^{p-\frac{1}{2}}$, 
where $p$ is a non-negative integer. 
For $d\geq 2$ and even, the singularities are logarithmic of the type:
$(\be-\be_0)^{p'}\ln^{q'}(\be-\be_0)$, 
where $p'$ is a non-negative integer and $q'$ a positive integer.
A slightly more complicated but essentially 
similar analysis applies for arbitrary anisotropies. The location of
the singularities will depend explicitly on the $\alpha_j$'s. 
These results can be made rigorous through the use of a Tauberian theorem.

The $\epsilon$ prescription becomes a convergence 
factor when one uses the series expansions (\ref{bessel})
of the Bessel functions. This gives
\beq
G^{(d)}_{\pm}(\vec{x}|\vec{\alpha},\be)= - \frac{1}{2\be^{1+X}} 
\sum_{n=0}^\infty c_{X+2n}^{(d)}(\vec{x}|\vec{\alpha}) \be^{-2n} \label{gres}
\eeq
where
\beq
c_n^{(d)}(\vec{x}|\vec{\alpha})\equiv
\int_{-\pi}^{+\pi}\cdots \int_{-\pi}^{+\pi}
\frac{d^d\vec{q}}{(2\pi)^d} \,\exp (i \vec{q}\cdot\vec{x})
\left(\sum_{j=1}^d \alpha_j\cos q_j \right)^n  \label{cn}
\eeq
and 
\bea
c_n^{(d)}(\vec{x}|\vec{\alpha}) &=& 0 \quad , \quad n=0,\cdots,X-1
\label{cn1} \\
c_{X+2n}^{(d)}(\vec{x}|\vec{\alpha}) &=& \left(X+2n\right)!\prod_{j=1}^d 
\left(\frac{\alpha_j}{2}\right)^{|x_j|}
\sum_{k_1\geq 0}\cdots \sum_{k_d\geq 0} 
\prod_{j=1}^d\left(\frac{\alpha_j}{2}\right)^{2 k_j}\label{cn2}\\
& &\times \frac{\delta_{k_1+\cdots +k_d,n}}{k_1 !\cdots k_d! (|x_1|+k_1)!
\cdots (|x_d|+k_d)!} \quad , \quad n\geq 0 \nonumber
\eea
In particular one finds 
\beq
G^{(d)}_{\pm}(\vec{x}|\vec{\alpha},\be) \sim 
-\frac{X!\,\prod_{j=1}^d\alpha_j^{|x_j|}}{|x_1|!
\cdots |x_d|!\, (2\be)^{1+X}}\quad\quad \quad
\be \rightarrow \infty
\eeq
which shows the Green function to vanish faster than the simple estimate 
$1/(2\be)$ obtained from the integral representation (\ref{gref}). 
A more compact form of the coefficients (\ref{cn2}) can be obtained 
by the pairwise replacement of Bessel functions through identity
(\ref{pairani}). This is done for $d=2,3,4$ in the following sections. 

The result (\ref{gres})  does not depend on the sign of the $\epsilon$
prescription. This is easily understood by noticing that 
for $|\be|>\sum_{j=1}^d |\alpha_j|$ the denominator in 
(\ref{gref}) does not have poles and therefore $\epsilon$ can be set to zero.
The large $\be$ expansion then yields (\ref{gres}) with the 
$c_n$ coefficients given by (\ref{cn}). 
One can also conclude that the expansion (\ref{gres}--\ref{cn2})
converges at least for $|\be|>\sum_{j=1}^d |\alpha_j|$. 
Convergence at the generalized 
massless point ($\be= \sum_{j=1}^d |\alpha_j|$) depends  
on the dimensionality of the lattice. This is related
to the recurrence/transience of the random walk. 
In one and two dimensions
the series diverge at this point, despite the $\epsilon$  prescription.
In  higher dimensions the series converge. 

The random walk interpretation of the $c_n$'s is the following. 
For a random walker starting from the origin, let $P_n(\vec{x})$ be
the probability of visiting  the site $\vec{x}$ after $n$ unit steps 
on the $d$-dimensional hypercubic lattice. Take the anisotropies 
to be positive and such that $\sum_{j=1}^d\alpha_j=d$. The probability
of jumping  from $\vec{x}$ to $\vec{x}+\hat{e}_j$ or to $\vec{x}-\hat{e}_j$
is  $\frac{\alpha_j}{2d}$. 
One then has: $P_n(\vec{x})=\frac{1}{d^n}\, c_n^{(d)}(\vec{x}|\vec{\alpha})$.
The vanishing of  $c_n$  for $n<X$ is therefore natural since 
$X$ is the minimal number of steps required to reach  point $\vec{x}$.

General  relations between the coefficients $c_n$ for different dimensions
can be simply found. Let $d'$ be any  positive integer smaller than $d$.
Expanding  $\sum_{j=1}^d\alpha_j \cos q_j$ 
using the binomial formula readily yields
\beq
c_n^{(d)}(\vec{x}|\vec{\alpha})=\sum_{k=0}^n {n \choose k}
c_k^{(d')}(x_1,\cdots,x_{d'}|\alpha_1,\cdots,\alpha_{d'})\, c_{n-k}^{(d-d')}(x_{d'+1},\cdots,x_d|\alpha_{d'+1},\cdots,\alpha_d)
\eeq
Note that these relations are valid for arbitrary anisotropies. 
One can also get other equations by expanding with the multinomial 
formul\ae. 

\section{The one-dimensional Green function}\label{d1}

In one dimension it is possible to obtain closed form expressions, 
and corresponding series expansions. The anisotropy parameter $\alpha_1$ 
is set to one  as it is an irrelevant overall factor. 

The closed forms are obtained by  integration in 
the complex $q$-plane.
The integration contour  is the rectangular path
$]-\pi +i\infty,-\pi] \cup [-\pi,\pi] \cup 
[\pi,\pi+i\infty[ \cup ]\pi+i\infty,-\pi+i\infty[$,
for the upper half-plane, and its reflection
about the real axis for the lower half-plane.
The part at infinity gives a vanishing contribution,
while the two side contributions cancel each other 
because $x\equiv x_1$ is integer. 
One  finds 
\beq
G^{(1)}_{\pm}(x|\be)= \pm \frac{e^{\pm i k |x|}}{2i\sin k} \quad , \quad
\be=\cos k\;\; , \quad  k\in {]0,\pi[} \label{d1gpm}
\eeq
One can compare (\ref{d1gpm}) to its continuum counter-part:
\beq
g^{(1)}_{\pm}(x|k)= \int_{-\infty}^{\infty} 
\frac{dq}{2\pi}\, \frac{\exp(iqx)}{k^2-q^2
\pm i\epsilon} =\pm \frac{e^{\pm i k |x|}}{2\,i\, k} \label{d1gpmc}
\eeq
One also finds 
\beq
\label{d1g}
G^{(1)}_+(x|\be)= G^{(1)}_-(x|\be) =\left\{
\begin{array}{rrr}
+\frac{e^{i k |x|}}{2i\sin k} 
\;\;,\quad  k_1\in {[0,\pi]}\;\;,\quad k_2 > 0  \\
\\
-\frac{e^{-i k |x|}}{2i\sin k} 
\;\;,\quad  k_1\in {]0,\pi[}\;\;,\quad k_2 < 0 
\end{array}\right.
\eeq
where $\be=\cos(k_1+ik_2)$ and $k=k_1+i k_2$.
Note that the positive exponential 
$\frac{e^{+k_2 |x|}}{2\sinh k_2}$ ($k_2 >0$),
despite satisfying equation (\ref{greq}),
is not obtained. This was to be expected from the 
integral representation (\ref{gref}), since for $|\be|>1$
the integrand has no singular point and the integral 
must vanish as $|x|\rightarrow \infty$.  

The general expression (\ref{gres}--\ref{cn2}) reduces to 
\bea
G^{(1)}_{\pm}(x|\be)&=& - \frac{1}{(2\be)^{|x|+1}}\sum_{k=0}^\infty 
\frac{(|x|+2k)!}{k! (|x|+k)!} \, (2\be)^{-2k}\label{g1}\\
 &=& -\frac{1}{(2\be)^{|x|+1}}\, _2F_1\left(\frac{|x|}{2}+\frac{1}{2}, 
\frac{|x|}{2}+1;|x|+1;\be^{-2}\right)\nonumber
\eea
This series converges uniformly for $|\be|>1$ and 
simply for $|\be|=1,\; \be\not=\pm 1$; it  diverges
at $\be=\pm 1$ and for $|\be|<1$.
The equality of this hypergeometric series  to the closed form 
(\ref{d1g}) was otherwise known \cite{bateman}.
The recurrence relation for the coefficients of the Green function,
with $c_n\equiv c_n^{(1)}(x|1)$, reads
\beq
(n^2-x^2)\, c_n - n(n-1)\, c_{n-2} =0 
\eeq 

The function $G^{(1)}_{\pm}$ satisfies a second-order 
differential equation in $\be$, of the hypergeometric type:
\beq
(\be^2 -1)\, y'' + 3\be\, y' + (1-x^2)\, y =0
\eeq
The indices at the 
three regular singular points  $\be=\pm 1$ and  $\infty$ 
are $(-\frac{1}{2},0)$  and $(1-|x|,1+|x|)$, respectively.
The two branch points $\be=\pm 1$ imply the existence of  
monodromies in the complex-mass plane. 
Solution (\ref{g1}) corresponds to $(\be=\infty\, ; \; s=1+|x|)$.
It is then natural to  investigate the properties of the solution
corresponding to $(\be=\infty\, ;\; s=1-|x|)$:
\bea
y_1(\be) &=& 2^{|x|-2} \be^{|x|-1} 
\, _2F_1\left(\frac{1-|x|}{2},1-\frac{|x|}{2};1-|x|;\be^{-2}
\right) \quad , \quad x\not= 0 \label{inf1mx}\\
y_{l1}(\be) &=& \frac{1}{4\be}\cdot \frac{1}{\sqrt{1-\be^{-2}}} 
\ln\left(\frac{1}{\be}\right)
+ \frac{1}{4}\sum_{n=0}^\infty\be^{-(2n+1)}\frac{d}{ds}
\left(\frac{\Gamma\left(n+\frac{s}{2}\right)\Gamma\left(\frac{s+1}{2}\right)}
{\Gamma\left(n+\frac{s+1}{2}\right)\Gamma\left(\frac{s}{2}\right)}
\right)_{|s=1} \label{inf1px}
\eea
with $_2F_1(0,\frac{1}{2};0,\be^{-2})\equiv 1$.
The derivative of the coefficients in (\ref{inf1px}) can be written 
using the  function $\psi(z)=\frac{d}{dz}\ln \Gamma(z)$.
The appearance of the logarithm for this second solution
is due the degeneracy of the indices at $x=0$. 
The hypergeometric series (\ref{inf1mx}) truncates to polynomials
in $\be^{+1}$. 
Note that this solution almost provides a Green function.
The {\it a priori} arbitrary normalization of a solution was
chosen to be equal to $2^{|x|-2}$ in (\ref{inf1mx}) so that
the one-dimensional Green equation,
$f(x+1|\be)+f(x-1|\be) -2\be\,  f(x|\be) = \delta_{x,0}$, 
is satisfied by $f(x|\be)$ given in (\ref{inf1mx}) and $f(0|\be)$ set to 0.

The expansion around $\be=1$ can be carried out similarly.
The two solutions are 
\bea
y_0(\be) &=& _2F_1\left(1-|x|,1+|x|;\frac{3}{2};\frac{1-\be}{2}\right)
\label{y0}\\
 &=& \frac{\sin\left(2|x|{\rm Arcsin}\left(
\sqrt{\frac{1-\be}{2}}\right)\right)}
{|x|\, \sqrt{1-\be^2}} \quad {\rm for} \quad \be \in ]-1,1]\label{y1}\\
y_{1/2}(\be) &=& \frac{1}{\sqrt{\be-1}} 
\, _2F_1\left(\frac{1}{2}-|x|,\frac{1}{2}
+|x|;\frac{1}{2};\frac{1-\be}{2}\right)\label{y2}\\
 &=&  -i \frac{\sqrt{2}\cos\left(2|x|{\rm
Arcsin}\left(\sqrt{\frac{1-\be}{2}}
\right)\right)}{\sqrt{1-\be^2}} \quad {\rm for} 
\quad \be \in ]-1,1[\label{y4}
\eea
These solutions are valid for all values of $x$.
In (\ref{y4}) the choice $\sqrt{-1}=i$ was made. 
The solution $y_0$ is regular at $\be=1$ and is a solution
to the Green equation when appropriately normalized, 
while $y_{1/2}$ is singular at $\be=1$, and is a solution
to the homogeneous  equation (\ref{greh}). 
Linear combinations of these solutions provide the two 
analytic continuations to the function defined by (\ref{g1}):
\beq
\frac{1}{2\sqrt{2}}\left(\sqrt{2} |x| y_0  \pm y_{1/2} \right)  =
\pm\frac{1}{2i\sin k} e^{\pm i k|x|}
\quad , \quad \be=\cos k \quad , \quad k\in ]0,\pi[ \label{lincomb1}
\eeq
Moreover one finds
\bea
& & \be^{|x|-1}\, _2F_1\left(\frac{1-|x|}{2},1-\frac{|x|}{2};1-|x|;\be^{-2}
\right) \nonumber\\ 
& & \phantom{+} =  2^{1-|x|}\, |x|\,
_2F_1\left(1-|x|,1+|x|;\frac{3}{2};\frac{1-\be}{2}\right) \quad , \quad
|x|\geq 1 
\eea
This corresponds to  the polynomial solutions 
(\ref{inf1mx}) and (\ref{y0}), which have to match
since, as polynomials, they are defined on the whole complex
$\be$-plane. 
(An amusing by-product of the foregoing analysis is the identity: 
$1=\sum_{n=1}^\infty \frac{(2n+1)!}{2^{3n}\, (2n-1)\,(n!)^2}\cdot$)
The expansion around $\be=-1$ yields similar results as 
can be expected from  parity. This symmetry is not
explicit on the series representations because 
the expansion point is not $\be=0$. 

The limits $\be\rightarrow \pm 1$ for the Green function
do not exist. However the following well-defined limit
\beq
G^{(1)}(x)\equiv\frac{1}{2}\lim_{\be\rightarrow 1^-}
\left( G^{(1)}_+(x|\be)+G^{(1)}_-(x|\be)\right) = \frac{1}{2} |x| \label{lim}
\eeq
is also  a solution  to the Green equation. 
In fact the latter equation is, for any $\be$, 
a one-dimensional recurrence relation 
which can be solved directly by the standard method. The points
$\be=\pm 1$ are  degeneracy points and the direct solution in this 
case is: $f(x)= (\pm 1)^x (f(0)\pm\frac{1}{2} |x|)$, 
for $\be=\pm 1$, and $f(0)$ arbitrary. 
For $\be=1$, and with the choice $f(0)=0$, one recovers (\ref{lim}).

\section{The two-dimensional Green function}\label{d2}

For arbitrary anisotropies,  
the identity (\ref{pairani}) for 
the  product of  two Bessel functions can be used to obtain
the following series expansion for the Green function:
\bea
G^{(2)}_{\pm}(\vec{x}|\vec{\alpha},\be) &=& -\frac{1}{|x_2|!\,
2\be^{X+1}}\cdot\left(\frac{\alpha_1}{2}\right)^{|x_1|}
\left(\frac{\alpha_2}{2}\right)^{|x_2|}\label{gres2a}\\
 & & \phantom{=} \times \sum_{k=0}^\infty \frac{(X+2k)!}{k!\; (|x_1|+k)! } \; _2F_1\left(-k,-|x_1|-k;|x_2|+1;\left(\frac{\alpha_2}{\alpha_1}\right)^2\right)
\cdot\left(\frac{\alpha_1}{2\be}\right)^{2k}\nonumber
\eea
When $\alpha_1=\alpha_2$, these anisotropies can be set to 1 and 
the preceding  expression becomes
\bea
G^{(2)}_{\pm}(\vec{x}|\be)&=& -\frac{1}{(2\be)^{X+1}}\, \sum_{k=0}^\infty
\frac{(X+k+1)_k (X+2k)!}{k!\,
(|x_1|+k)! \, (|x_2|+k)!} (2\be)^{-2k} \label{gres2}\\
 &=& -\frac{X!}{(2\be)^{X+1}\,|x_1|! \, |x_2|!}  \\
 & & \times \, _4F_3 (\frac{X+1}{2},\frac{X+1}{2},\frac{X}{2}+1;
\frac{X}{2}+1; X+1,|x_1|+1,|x_2|+1;4\be^{-2})\nonumber
\eea
The series (\ref{gres2}) converges uniformly  for 
$|\be|> 2$, simply for $|\be|=2$ and $\be\not=\pm 2$ and
diverges at $\be=\pm 2$ and $|\be|<2$. 

The recurrence relation for the coefficients of the Green function read
\beq
(n^2 -X^2) (n^2 - x^2) \, c_n - 4 n^2(n-1)^2 c_{n-2} =0 \label{rec2inf}
\eeq
where $x\equiv |x_1|- |x_2|$.
The Green function $G^{(2)}_{\pm}$ satisfies a $4^{\rm th}$ order
differential equation in $\be$:
\bea
& & \be^2 (\be^2 -4)\, y'''' + \be\, (10\be^2 -16)\, y''' 
+ \left(\be^2 (-2(x_1^2+x_2^2)+25)-8\right) y'' \nonumber\\
& & \phantom{+} +3\be \,\left( -2(x_1^2+x_2^2)+5\right) y'
+(1- X^2)(1- x^2) \, y = 0
\eea
The four regular singular points  and their indices are:
\beas
\be &=& 0\; : \quad s= 1,1,0,0\\
\be &=& \pm 2\; : \quad s= 2,1,0,0\\
\be &=&\infty\; : \quad s= 1+X
\, ,\; 1-X\,,\;1+x\,,\; 1-x 
\eeas
The three finite singular points were predicted in section 2. 
The series (\ref{gres2}) is the regular solution corresponding to 
$(\be=\infty\, ;\; s=1+X)$. 
The degeneracy of the indices at the other points signals the 
existence of logarithmic solutions. 

The radius of convergence of the series expansion around $\be=0$ 
is equal to 2, as $\be=\pm 2$ are the closest singularities. 
Thus this  expansion and  the expansion around infinity
cover the whole complex $\be$-plane.
The recurrence relation for the coefficients of the series expansions 
around $\be=0$ contains two terms, just like (\ref{rec2inf}) for the 
expansion around infinity. This permits the explicit determination
of the  series. One also has to find the connection 
coefficients involved in the linear combinations of 
the four solutions which reproduce the 
Green functions at hand, or the analytic continuations 
of (\ref{gres2}). In the previous section, in (\ref{lincomb1}), one had
$\frac{|x|}{2}$ and $\pm \frac{1}{2\sqrt{2}}$. 
But the determination of the appropriate functions of $\vec{x}$ is 
not an easy task and there is no  general 
systematic method which gives  a closed form result. 
Here it is possible to carry out this analysis completely, 
and I have found the following expansions and connection coefficients:
\beq
G^{(2)}_{\pm}(\vec{x}|\be) =  [f_0(\vec{x}) \pm l_0(\vec{x})]\,
y_0 (\be) + [f_1(\vec{x}) \pm l_1(\vec{x})]\, y_1 (\be)
\pm h_0(\vec{x})\,  y_{l0}(\be) \pm h_1(\vec{x}) \, y_{l1}(\be)\label{g20}
\eeq
where
\bea
& & y_0 (\be) = \, _4F_3 \left(\frac{1+X}{2},\frac{1-X}{2},\frac{1+x}{2};
\frac{1-x}{2};1,\frac{1}{2},\frac{1}{2};\frac{\be^2}{4}\right)\\
& & y_1 (\be) = \be \;  _4F_3 \left(\frac{2+X}{2},\frac{2-X}{2},\frac{2+x}{2};
\frac{2-x}{2};\frac{3}{2},\frac{3}{2},1;\frac{\be^2}{4}\right)\\
& & y_{l0} (\be) = \, y_0 (\be) \ln\be + \frac{d}{ds}\,
_5F_4 \left(\frac{s+1+X}{2},\frac{s+1-X}{2},\frac{s+1+x}{2};
\frac{s+1-x}{2},1; \right.\\
& & \phantom{y_0 (\be) =} \left. \frac{s+2}{2},\frac{s+2}{2},
\frac{s+1}{2},\frac{s+1}{2};\frac{\be^2}{4}\right)_{|s=0}\nonumber\\
& & y_{l1} (\be) = \, y_1 (\be) \ln\be + \be \;
\frac{d}{ds} \, _5F_4 \left(\frac{s+1+X}{2},\frac{s+1-X}{2},\frac{s+1+x}{2};
\frac{s+1-x}{2},1;\right. \\
& & \phantom{y_0 (\be) =} \left. \frac{s+2}{2},\frac{s+2}{2},
\frac{s+1}{2},\frac{s+1}{2};\frac{\be^2}{4}\right)_{|s=1}\nonumber
\eea
are the four independent solutions of the differential equation 
around $\be=0$. The $s$-derivatives  can 
be easily expressed in terms of $\psi(z)$, the logarithmic derivative of
the Gamma function. The logarithms are taken real for positive $\be$. 
The six connection functions are given by:
\bea
f_0(\vec{x}) &=& \frac{1}{4} \cos\left(\frac{\pi}{2}X\right)
\cos\left(\frac{\pi}{2}x\right)
+ \frac{1}{4}\sin\left(\frac{\pi}{2}X\right) 
\sin\left(\frac{\pi}{2}|x|\right)\\
f_1(\vec{x}) &=& \frac{1}{2} |x_1^2-x_2^2| \, f_0(\vec{x}) 
= \frac{1}{2}\, |x|\, X \, f_0(\vec{x})\\
h_0(\vec{x}) &=& h_0 \cos\left(\frac{\pi}{2}X\right) \cos\left(
\frac{\pi}{2}x\right)\\
h_1(\vec{x}) &=& \frac{h_0}{2}|x_1^2-x_2^2| \,
\sin\left(\frac{\pi}{2}X\right) \sin\left(\frac{\pi}{2}|x|\right) = 
\frac{h_0}{2}\, x\, X \,\sin\left(\frac{\pi}{2}X\right) 
\sin\left(\frac{\pi}{2}x\right)\\
l_0(\vec{x}) &=& \left( \psi\left(\frac{X+1}{2}\right)  + 
\psi\left(\frac{|x|+1}{2}\right)\right) \, h_0(\vec{x})\\
l_1(\vec{x}) &=& \left( \psi\left(\frac{X}{2}+1\right)  + 
\psi\left(\frac{|x|}{2}+1\right) -2 \right) \, h_1(\vec{x})\\
\phantom{l_1(\vec{x})} & &
-\frac{h_0}{2} (|x|+X) \sin\left(\frac{\pi}{2}X\right) 
\sin\left(\frac{\pi}{2}|x|\right) \nonumber
\eea
where $h_0= \frac{i}{2\pi}\cdot$ 
These functions satisfy a number of equations of
which the simplest are:
\bea
& & H f_0(\vec{x}) = \delta_{\vec{x},\vec{0}}\; ,\quad H h_0(\vec{x}) = 0
\; , \quad H l_0(\vec{x}) = 0 \\
& & H f_1(\vec{x}) = 2 f_0(\vec{x}) \; , \quad 
H h_1(\vec{x}) = 2 h_0(\vec{x}) \; , \quad H l_1(\vec{x}) = 2 l_0(\vec{x})\\
& & x_1 [ f_0(x_1 + 1,x_2) - f_0(x_1-1,x_2) ] =
x_2[ f_0(x_1,x_2+1) - f_0(x_1,x_2-1) ] 
\eea
Thus $f_0$ is a Green function at the decoupling point
while $h_0$ and $l_0$ are solutions of the homogeneous equation 
at the same point. The functions $f_1$, $h_1$ and $l_1$ appear 
as potentials for the sources  $f_0$, $h_0$ and $l_0$, respectively.
The last equation for $f_0$ is a strong form of a 
directional-independence relation.
The $f_0-f_1$ part in (\ref{g20}) is a solution, regular at $\be=0$, 
of the Green equation. The remaining part is a singular solution 
of the homogeneous equation, independently from the value of $h_0$. 
Finally, $h_0(\vec{x})y_0(\be) + h_1(\vec{x}) y_1(\be)$ is 
a regular solution of the homogeneous equation. 

The parity property is satisfied with $\ln(-1)= \pm i\pi$,
and the complex conjugation symmetry for both $\be$ and $\be^*$ 
not on the branch cut. 

Consider now the expansion about $\be=2$, and let $v=\be-2$. 
The index $s$ can take one of the three values already found: $0$, $1$ or 2. 
A  generic solution is given by
\beq
y(v) = \sum_{n\geq 0} a_n(s)\, v^{n+s}
\eeq
where
\bea
& & 16 (n+s)^2 (n+s-1)(n+s-2) \, a_n \\
& & \phantom{+} + 4  (n+s-1)(n+s-2)\left[5 n^2 +(10 s-9) n +5 +5 s^2 -9 s  
-2(x_1^2+x_2^2)\right] \, a_{n-1}\nonumber\\
& & \phantom{+} +  2 (2n+2s-3)(n+s-2)\left[2 n^2 +(4 s-6) n +5 +2 s^2 -6 s  
-2(x_1^2+x_2^2)\right] \, a_{n-2}\nonumber\\
& & \phantom{+} +   (n+s-2+X) 
(n+s-2-X)(n+s-2+x)(n+s-2-x)\, a_{n-3} =0 \nonumber
\eea
with $n\geq 0$ and $a_{<0}\equiv 0$.
For $a_0(s)\equiv 1$, the solutions
\bea
& & y_2(v) = \sum_{n\geq 0} a_n(s=2) \, v^{n+2} \\
& & y_1(v) = \sum_{n\geq 0} a_n(s=1) \, v^{n+1} \; ,\quad \quad a_1 \equiv 0\\
& & y_0(v) = \sum_{n\geq 0} a_n(s=0) \, v^n \; , \quad\quad a_1=a_2 \equiv 0
\eea
are linearly independent and regular at $\be=2$. 
The fourth solution has the expected logarithmic singularity:
\beq
y_{l0}(v)= y_0(v)\, \ln v + \sum_{n\geq 0}v^n  \frac{d}{ds}a_n(s)_{|s=0} 
\eeq
The complete Green function is {\it a priori} a linear
combination of the four solutions. I have not looked for the 
corresponding connection coefficients. However one can conclude
that the logarithmic solution corresponds to a solution of 
the homogeneous equation (\ref{greh}), and can therefore be dropped
without altering the Green property. The remaining piece
is the natural regularization at $\be=2$ of the Green function
defined by (\ref{gref}). The situation at $\be=-2$ is similar. 

For $x_1=x_2$ and arbitrary anisotropies, 
Montroll found an expression in terms of Legendre functions \cite{montroll}.
In terms of the definitions adopted here one has: 
\beq
G^{(2)}_{\rm M}((x_1,x_1)|(\alpha_1,\alpha_2),\be)= 
-\frac{1}{2\pi (\alpha_1\alpha_2)^{\frac{1}{2}}}\, Q_{|x_1|-\frac{1}{2}}
\left(\frac{\be^2-(\alpha_1^2+\alpha_2^2)}{2\alpha_1\alpha_2}\right)
\label{gmont}
\eeq
where $Q_{\nu}(z)$ is the $\nu^{\rm th}$ Legendre function of the second kind. 
These functions have logarithmic singularities at $z=\pm 1$. But 
for $\nu$ a half odd-integer, only $z=1$ is a singularity:
\beq
Q_{|x_1|-\frac{1}{2}}(1-\epsilon) \sim -\frac{1}{2} \ln\left(\frac{\epsilon}{2}
\right) -\gamma -\psi\left(|x_1|+\frac{1}{2}\right) + O(\epsilon) \; , 
\quad \quad \epsilon \rightarrow 0^+
\eeq
where $\gamma=0.577216\cdots$. 
Thus one finds a logarithmic divergence at $\be=\pm 2$:
\bea
& & G^{(2)}_{\rm M}((x_1,x_1)|(1,1),\be)= 
\frac{1}{4\pi} \ln\left(\frac{4-\be^2 }{4}\right)\\ 
& & \phantom{G^{(2)}_{\pm}} +\frac{\gamma}{2\pi}
+\frac{1}{2\pi}\psi\left(|x_1|+\frac{1}{2}\right) 
+ O\left(\frac{\be^2-4}{2}\right) \; ,\quad \quad
\be\rightarrow 2^- \; {\rm or}\; \be \rightarrow -2^+ \nonumber
\eea
One also has $G^{(2)}_{\rm M}((x_1,x_1)|(1,1),0)= -\frac{1}{4}\cdot$

Montroll's result should be qualified. It is general in terms of 
anisotropies, but partial as it applies only to 
the line $x_1=x_2$ and $|\be|<2$. Also,  it  is only a part of the full 
Green function, even under these restrictions.  
Consider $x_1=x_2$ in  the Green function given by (\ref{g20}). 
At $\be=0$ this expression diverges, as expected from 
the analysis of section 2,
and thus  contradicts (\ref{gmont}). However taking the
half-sum of the two Green functions one finds:
\beq
G^{(2)}(\vec{x}|\be) =  \frac{1}{2}\left( 
G^{(2)}_+(\vec{x}|\be)+G^{(2)}_-(\vec{x}|\be) \right) = 
f_0(\vec{x})\, y_0 (\be) + f_1(\vec{x}) \, y_1 (\be)\label{g20r}
\eeq
This solution of the Green equation is regular at $\be=0$, 
and
\beq
G^{(2)}((x_1,x_1)|\be) = \frac{1}{4}\cos(\pi x_1)\, _2F_1\left(\frac{1}{2}+|x_1|,\frac{1}{2}-|x_1|;1;
\frac{b^2}{4}\right)
\eeq
For $x_1=0$, one has the following identity:
\beq
_2F_1\left(\frac{1}{2},\frac{1}{2};1;\frac{b^2}{4}\right)=\frac{2}{\pi} 
K\left(\frac{\be^2}{4}\right)
\eeq
where $K$ is the complete elliptic function of the first kind \cite{abraste}.
One also has  
\beq
Q_{|x_1|-\frac{1}{2}}\left(\frac{\be^2-2}{2}\right)= \frac{\pi}{2}\cos(\pi x_1)
\, _2F_1\left(\frac{1}{2}+|x_1|,\frac{1}{2}-|x_1|;1;
\frac{b^2}{4}\right)\; , \quad\quad |\be|<2 
\eeq
which shows that $G^{(2)}((x_1,x_1)|\be)$ and 
$G^{(2)}_{\rm M}((x_1,x_1)|(1,1),\be)$ are equal, up to a factor of $-1$. 
The origin of this sign in (\ref{gmont}) is unclear.
Compare now (\ref{gmont}) to the half-sum of the two functions 
(\ref{gres2}) at  $x_1=x_2$.
The former function is even in $\be$ while the latter is odd. 
Therefore they cannot be equal, and (\ref{gmont}) holds only for $|\be|< 2$.

\section{The three-dimensional Green function}\label{d3}

Using the identity (\ref{pairani}) for 
the  product of  two Bessel functions one obtains
the following series expansion for the three-dimensional function:
\bea
G^{(3)}_{\pm}(\vec{x}|\vec{\alpha},\be) &=& -\frac{1}{|x_2|!\;
2 \be^{X+1}}\cdot\left(\frac{\alpha_1}{2}\right)^{|x_1|}
\left(\frac{\alpha_2}{2}\right)^{|x_2|}
\left(\frac{\alpha_3}{2}\right)^{|x_3|}\label{gres3a}\\
 & & \phantom{=} \times  \sum_{k=0}^\infty (X+2k)!
\left(\frac{\alpha_3}{2\be}\right)^{2k}\sum_{p=0}^k
\frac{1}{p!\, (k-p)!\, (|x_1|+p)!\, (|x_3|+k-p)!} \nonumber\\
 & & \phantom{=} \times 
\left(\frac{\alpha_1}{\alpha_3}\right)^{2p}\;
_2F_1\left(-p,-|x_1|-p;|x_2|+1;\left(\frac{\alpha_2}
{\alpha_1}\right)^2\right)\nonumber
\eea
When the anisotropies $\alpha_j$  are equal to 1  
the preceding  expression simplifies to 
\bea
G^{(3)}_{\pm}(\vec{x}|\be) &=& -\frac{1}{(2\be)^{X+1}}\, \sum_{k=0}^\infty
\frac{(X+2 k)!}{(2\be)^{2k}}\label{gres3}\\
 & & \phantom{=} \times\sum_{p=0}^k\frac{(|x_1|+|x_2|+p+1)_p}{p!\, (k-p)!\,(|x_1|+p)! \, (|x_2|+p)! \, (|x_3|+k-p)!}\nonumber
\eea
The series (\ref{gres3}) converges for $|\be|\geq 3$.
For  $\vec{x}=\vec{0}$ one can write 
\bea
G^{(3)}_{\pm}(\vec{0}|\be) &=& -\frac{1}{2\be}\sum_{k=0}^\infty 
\frac{(2k)!}{4^{k} \, (k!)^2}\, u_k \,\be^{-2k}\label{ser3}\\
u_k&=&\sum_{p=0}^k {k\choose p}^2 {2p \choose p} 
\eea
The value of this series  at $\be=3$  
was calculated by Watson \cite{watson}:
\beq 
G^{(3)}_{\pm}(\vec{0}|3)=
-\frac{2}{\pi^2}\left[18+12\sqrt{2}-10\sqrt{3}-7\sqrt{6}\right]
K^2\left((2-\sqrt{3})^2(\sqrt{3}-\sqrt{2})^2\right) \label{watsy}
\eeq
The numerical value of (\ref{watsy}) is: $-0.2527\cdots$
(The definition of \cite{abraste} is adopted for the function $K$.)
At the singular point $\be=3$, this series converges, rather slowly,
to the known value (\ref{watsy}).
The sum of the first 1001 terms gives: $-0.2502\cdots$. 
It is amusing to note that the  $u_k$ ($k\geq 1$)
appear to be divisible by 3, the number of dimensions. 

It is possible to derive a  recurrence relation 
for the coefficients of the $3$-dimensional Green function. 
Define the following even homogeneous polynomials
\beas
& & \Sigma_{222} = x_1^2 x_2^2 x_3^2 \\
& & \Sigma_{422} = x_1^2 x_2^2 x_3^2 \, ( x_1^2 + x_2^2 + x_3^2 )\\
& & \Sigma_{2i} = x_1^{2i}+x_2^{2i} +x_3^{2i} \; , \quad i=1,2,3,4\\
& & \Sigma_{(2i)(2i)} = x_1^{2i}x_2^{2i}+x_2^{2i}x_3^{2i} 
+x_1^{2i}x_3^{2i} \; , \quad i=1,2\\
& & \Sigma_{(2i)2} = x_1^{2i} x_2^2 + x_1^2 x_2^{2i} + x_2^{2i} x_3^2 + x_2^2 
x_3^{2i} + x_1^{2i} x_3^2 + x_1^2 x_3^{2i} \; , \quad i=2,3
\eeas
For $c_n\equiv c_n^{(3)}(\vec{x}|(1,\cdots,1))$, I have found 
\bea
& & (n-2)(n-4)\left[ \Sigma_8 -4 \Sigma_{62} + 6 \Sigma_{44} +4 \Sigma_{422}
\phantom{n^8}\right. \\
& & \left. -4 n^2 \left( \Sigma_6 -\Sigma_{42} +10 \Sigma_{222}\right)+ 2 n^4
\left(3\Sigma_4+ 2 \Sigma_{22}\right) 
-4 n^6\Sigma_2 +n^8 \right] c_n \nonumber\\
& & +4 n (n-1)^2 (n-4)\left[-\Sigma_6 +\Sigma_{42} + 6 \Sigma_{222}
-(n^2-4 n-2)\Sigma_4 -2 (3 n^2 -4 n+2)\Sigma_{22} \right.\nonumber\\
& & \left. + n(n-2)(5 n^2 -6 n+8)
\Sigma_2 - n (n-2)(n^2 +2)(3 n^2 -6  n +4)\right] c_{n-2}\nonumber\\
& & + 2 n(n-1)(n-2)(n-3)\left[-(n^2-4 n+12)\Sigma_4 
+ 2 (5 n^2 -20 n +12)\Sigma_{22} \right. \nonumber\\
& & \left. - 2n (n-4) (7 n^2 -24 n +28)\Sigma_2
+ n(n-4) (15 n^4 -96 n^3 + 268 n^2 -384 n +248 ) \right] c_{n-4}\nonumber\\
& & +4 n^2 (n-1) (n-2)(n-3)(n-4)(n-5)\left[ 3 (n-3) \Sigma_2 - (7 n^3 -57 n^2
+158 n -162)\right] c_{n-6} \nonumber\\
& & + 9 n^2 (n-1)(n-2)^2(n-3)(n-4)(n-5)(n-6)(n-7) \, c_{n-8} = 0\nonumber
\eea
This translates into a $10^{\rm th}$ order differential equation 
for $y=G^{(3)}_{\pm}$: 
\bea
& & \be^2 (\be^2 -1)^3 (\be^2 -9)\, y^{(10)} + 
\be\,(\be^2-1)^2 (61 \be^4 - 418 \be^2 + 45)\, y^{(9)} \\
& & -(\be^2-1)\left[ \be^6 ( 4 \Sigma_2 - 1433) + \be^4 (- 16\Sigma_2 + 7511)
+ \be^2 (12 \Sigma_2  - 2673 ) + 27 \right]\, y^{(8)} \nonumber\\
& & -4 \be \,\left[\be^6 (42\Sigma_2 - 4167) + \be^4( -146\Sigma_2 + 17284)
 + \be^2( 122\Sigma_2 - 11683)  - 18\Sigma_2 + 1140 \right] y^{(7)} \nonumber\\
& & +\left[ \be^6 (6\Sigma_4 + 4 \Sigma_{22} - 2552 \Sigma_2 + 102963)  
+ \be^4(- 4 \Sigma_4 - 24 \Sigma_{22} 
+ 5740 \Sigma_2 - 261972)\right.\nonumber\\
& & \left. + \be^2( - 2\Sigma_4   + 20 \Sigma_{22} - 2444 \Sigma_2 + 90750 )   
+ 48\Sigma_2  -1536\right] y^{(6)} \nonumber\\
& & - \be \left[\be^4 (- 162 \Sigma_4 - 108 \Sigma_{22} 
+ 17640 \Sigma_2 - 337617)\right. \nonumber \\
& & \left. + \be^2 (76 \Sigma_4  + 392 \Sigma_{22} - 23180 \Sigma_2 + 470364) 
+ 14\Sigma_4  - 140\Sigma_{22} + 3812\Sigma_2 - 64194 
\right] y^{(5)} \nonumber \\
& &  -2 \left[ \be^4( 2 \Sigma_6 -  2 \Sigma_{42} + 20 \Sigma_{222} 
- 714 \Sigma_4 - 476 \Sigma_{22} + 28742 \Sigma_2  - 278744 )\right.\nonumber\\
& & + \be^2( 2 \Sigma_6 - 2 \Sigma_{42} - 12 \Sigma_{222} + 202 \Sigma_4  
+ 892 \Sigma_{22} \nonumber \\
& & \left. - 18870 \Sigma_2^{\phantom 4} + 179166 ) 
+ 17\Sigma_4  - 74\Sigma_{22} + 590\Sigma_2 - 5055 \right] \,y^{(4)}\nonumber\\
& & -4 \be\,\left[\be^2(16 \Sigma_6  - 16 \Sigma_{42} + 160 \Sigma_{222} 
- 1230 \Sigma_4 - 820 \Sigma_{22} + 20776 \Sigma_2 - 103135 )\right.\nonumber \\
& & \left. + 8\Sigma_6^{\phantom 4} - 8\Sigma_{42} -  48\Sigma_{222}  
+ 150\Sigma_4  + 612\Sigma_{22} - 5190\Sigma_2  
+ 23022\right]\,y^{(3)}\nonumber\\
& & +\left[ \be^2(\Sigma_8 - 4 \Sigma_{62} + 6 \Sigma_{44} + 4 \Sigma_{422} 
-  276 \Sigma_6 + 276 \Sigma_{42} - 2760 \Sigma_{222} \right. \nonumber\\
& & + 6246 \Sigma_4  + 4164 \Sigma_{22} - 44916 \Sigma_2 + 108681) \nonumber\\
& & - 40\Sigma_6 +  40\Sigma_{42} + 240\Sigma_{222}
\left. - 120\Sigma_4^{\phantom 4}  - 720\Sigma_{22} 
+ 2280\Sigma_2  -4680 \right] \, y^{(2)} \nonumber\\
& & +9 \be\,\left[ \Sigma_8^{\phantom 4} 
- 4\Sigma_{62} + 6\Sigma_{44} +  4\Sigma_{422} 
- 36\Sigma_6  \right. \nonumber\\
& & \left. + 36\Sigma_{42} - 360\Sigma_{222} 
+ 246\Sigma_4^{\phantom 4} + 164\Sigma_{22} 
- 676\Sigma_2 + 681)\right] \, y^{(1)}\nonumber\\
& & + 15\,\left[\Sigma_8^{\phantom 4} 
-  4\Sigma_{62}  + 6\Sigma_{44} + 4\Sigma_{422}
- 4\Sigma_6 + 4\Sigma_{42} - 40\Sigma_{222} 
+ 6\Sigma_4 + 4\Sigma_{22} - 4\Sigma_2 + 1 \right] \,  y = 0 \nonumber
\eea
This equation has six regular 
singular points, 0, $\pm 1$, $\pm 3$ and $\infty$.
The corresponding indices are:
\beas
\be &=& 0\; : \quad s= 7,7,6,5,5,4,3,2,1,0\\
\be &=& \pm 1\; : \quad s= 6,5,4,3,2,1,0,\;
\frac{5}{2}\raisebox{0.5ex}{,} \frac{3}{2}\raisebox{0.5ex}{,}\frac{1}{2}\\
\be &=& \pm 3\; : \quad s= 8,7,6,5,4,3,2,1,0,\;\frac{1}{2}\\
\be &=&\infty\; : \quad s= 5,3,\; 1+|x_1|+|x_2|+|x_3|
\, ,\; 1-|x_1|-|x_2|-|x_3|\,,\\
\phantom{\be} &\phantom{=}& \phantom{jjjjjjjjjj}
1+|x_1|+|x_2|-|x_3|\,,\; 1+|x_1|-|x_2|+|x_3|\,,\; 1-|x_1|+|x_2|+|x_3|\,,\\
\phantom{\be} &\phantom{=}& \phantom{jjjjjjjjjj}
1+|x_1|-|x_2|-|x_3|\,,\; 1-|x_1|-|x_2|+|x_3|\,,\; 1-|x_1|+|x_2|-|x_3| 
\eeas
The Green function is regular at  $\be=\infty$ and corresponds to the 
index $1+|x_1|+|x_2|+|x_3|$. 
The appearance of indices differing by integer values can result
in logarithms. However the 10-term recurrence for the series expansion
around $\be=\pm 3$ shows that all 10 solutions do not contain
logarithms. The same conclusion holds at  $\be=\pm 1$, with an 8-term
recurrence. 
Therefore  $\be=\pm 1$ and $\be=\pm 3$ are branch point singularities
which are free of logarithms. 
At $\be=0$ the indices indicate that some solutions contain logarithms. 
For the foregoing Green function  $\be=0$ is a regular point,  
and one should consider a linear combination
of the logarithm-free solutions. 

The reason for the appearance of $\be=0$ as a singularity 
of the differential equation is unclear. Perhaps the fact that this 
point is a fixed point of the parity symmetry, or its status as a decoupling
point may be relevant here. 

At its four singular points, the Green function does not diverge 
as all the indices are non-negative;  only the solutions which correspond
to the vanishing indices give non-vanishing contributions. 
However, around a given singular point, one can  still consider a natural regularization by dropping all the solutions in the corresponding
linear combination which are associated with non-integer indices. 
Such solutions combine into  a solution of the homogeneous equation
(\ref{greh}). 
Finally  note that some results 
were obtained by Joyce \cite{joyce}, 
at $\vec{x}=\vec{0}$ and around $\be=3$. 

\section{Concluding remarks}\label{d4}

Using  equation (\ref{pairani}) twice gives the large-$\be$ series
expansion of the four-dimensional Green function:
\bea
G^{(4)}_{\pm}(\vec{x}|\vec{\alpha},\be) &=& -\frac{\prod_{j=1}^4 \left(\frac{\alpha_j}{2}\right)^{|x_j|}}{|x_2|!\,|x_4|!}\cdot
\frac{1}{2\be^{X+1}}\label{gres4a} \sum_{k=0}^\infty (X+2k)! 
\left(\frac{\alpha_3}{2\be}\right)^{2k}\\
 & & \phantom{=} \times \sum_{p=0}^k  \;
\frac{1}{p!\,(k-p)!\, (|x_1|+p)!\,(|x_3|+k-p)!} \left(\frac{\alpha_1}{\alpha_3}
\right)^{2p} \nonumber\\ 
& & \phantom{=} \times \;
_2F_1\left(-p,-|x_1|-p;|x_2|+1;\left(\frac{\alpha_2}{\alpha_1}\right)^2\right)
\nonumber\\
& & \phantom{=} \times \;
_2F_1\left(-(k-p),-|x_3|-(k-p);|x_4|+1;
\left(\frac{\alpha_4}{\alpha_3}\right)^2\right)\nonumber
\eea
When the anisotropies are all equal to 1  
the preceding  expression becomes 
\bea
G^{(4)}_{\pm}(\vec{x}|\be) &=& -\frac{1}{(2\be)^{(X+1)}}\, \sum_{k=0}^\infty
\frac{(X+2k)!}{(2\be)^{2k}}\label{gres4}\\
& & \phantom{=} \times  \sum_{p=0}^k \;
\frac{(|x_1|+|x_2|+p+1)_p (|x_3|+|x_4|+k-p+1)_{k-p}}{p!\,(k-p)! \,
(|x_1|+p)! \, (|x_2|+p)! \,(|x_3|+k-p)!\,(|x_4|+k-p)!}\nonumber
\eea
The series (\ref{gres4}) converges for $|\be|\geq 4$.
For  $\vec{x}=\vec{0}$ one can write 
\bea
G^{(4)}_{\pm}(\vec{0}|\be) &=& -\frac{1}{2\be}\sum_{k=0}^\infty 
\frac{(2k)!}{4^{k} \, (k!)^2}\, v_k \,\be^{-2k}\\
v_k&=&\sum_{p=0}^k {k\choose p}^2 {2k-2p\choose k-p}{2p \choose p}
\eea
For $\be=4$, this series converges to the known value of $-0.1549\cdots$
\cite{luwe2}.
The sum of the first 31 terms gives: $-0.1541\cdots$. 
Similarly to the $u_k$ in the preceding section, the $v_k$ ($k\geq 1$)
also appear to be divisible by the number of dimensions, here equal to 4. 

A derivation of the recurrence relation for the coefficients
$c_n^{(d)}(\vec{x})$ can be done as for the lower dimensions. 
The order of the recurrence appears to be larger than 7. From this 
recurrence a differential equation can be derived. 
The 5 singular points are of the logarithmic type. The logarithmic
solutions, at one given singular point,  
can be dropped leaving a regular Green function. 

These features were seen to be common to the lowest dimensions.
They also hold for all the higher dimensions. 
The coefficients $c_n^{(d)}(\vec{x})$ satisfy recurrence
relations for all dimensions. The general form of these relations
is easily inferred from the results for the lower dimensions. 
The coefficients appearing in the recurrence relations are 
polynomials in $n$ and the $x_j^2$'s. From such relations one can then
derive the differential equation as was done 
for the lowest dimensions. 
Another common feature is the possibility of dropping the singular
part, around one  given singularity. This part is a solution of the 
homogeneous equation.  
Finally, at $\be=d$, subtracting $G_{\pm}^{(d)}(\vec{0}|d)$ from (\ref{gref})
provides another regularization (see \cite{shin} for $d=2$). 

Determining the explicit recurrence relation and the differential equation 
is however not a trivial task. It is also difficult to find  
the $\vec{x}$-dependent coefficients appearing
in the linear combination of the solutions around 
a given singularity. These techniques were applied for the lower dimensions
and new results were obtained. 
While the low dimensions studied in this paper
seem to be at the limit of tractability of these methods,  
the knowledge obtained about the analytic structure of the lattice Green 
functions in all dimensions is an important step in their study.
\\

\noindent{\bf Acknowledgements:} I would like to thank
Ivan Horv\'ath for bringing to my attention reference \cite{luwe2} 
and for a critical reading of the manuscript.   
Special thanks to Tim Newman for bringing to my 
attention reference \cite{hughes}, for a critical reading of the manuscript,
for his lightning derivation of $\frac{i}{2\pi}$   
and for enjoyable  coffee breaks. 
I thank P. Arnold, P. Fendley and H. Thacker for discussions, and 
S. Adhikari and G. Moore    for communications.  
\\

\noindent{\Large \bf Appendix A: Bessel functions  and other formul\ae}
\\

\noindent The  cylindrical  Bessel functions $J_n(z)$
have both integral representations 
\beq
J_n (z)=\frac{i^{-n}}{\pi} \int_0^{\pi}dq \exp(iz \cos q) \cos(nq)\;\; , \;\;
\; \forall \, n \, \in \bZ \label{intrep}
\eeq
and series expansions
\beq
J_n(z) = \left(\frac{z}{2}\right)^n\sum_{k=0}^\infty 
\frac{(-1)^k}{k!\,\Gamma(n+k+1)}\, \left(\frac{z}{2}
\right)^{2k}\; ,\label{bessel}
\eeq
with an infinite radius of convergence.
The asymptotic behavior at infinity is given by
\beq
J_n(z) \sim \sqrt{\frac{2}{\pi z}}\left(\cos\chi -\frac{4n^2-1}{8z}\sin\chi 
\right)\;\; , \quad |z|
\rightarrow \infty \;\; , \quad |{\rm arg}(z)|<\pi \label{assbes}
\eeq
where $\chi=z-\frac{\pi}{2} n -\frac{\pi}{4}\cdot$  One also has
\beq
\int_0^\infty J_n(x)\,dx = 1 \;\; ,\quad n \geq 0
\eeq
The Bessel functions  have the following properties
\bea
& & J_n(-z)=(-1)^n J_n(z) \;\; \; \forall \, n \; \in \bZ \label{prop1}\\
& & J_{-n}(z)=(-1)^n J_n(z) \;\;\;\; \;\;\;\forall \, n \; 
\in \bZ \label{prop2}\\
& & J_0(0)=+1 \;\;,\;\;\;\; J_n(0)= 0 \;\;, \;\;\; n 
\not= 0\label{prop3}\\
& & J_{n-1}(z) - J_{n+1}(z) = 2 J'_n(z) \;\;\; \forall \, n \, 
\in \bZ \label{prop4}
\eea
A particular formula for the product of two Bessel functions is 
\beq
J_m (az) J_n (bz)=\frac{\left(\frac{az}{2}\right)^m 
\left(\frac{bz}{2}\right)^n}{\Gamma(n+1)} \sum_{k=0}^\infty \frac{(-1)^k\ 
_2F_1(-k,-m-k;n+1;\frac{b^2}{a^2})}{k!\,\, \Gamma(m+k+1)} 
\left(\frac{az}{2}\right)^{2k} \label{pairani}
\eeq
(A typographical error in \cite{gradz} has been corrected.)
Note that $_2F_1$ is in fact a polynomial in $b^2/a^2$. 
When $a=b=1$ this formula simplifies to 
\beq
J_m (z) J_n (z)= \sum_{k=0}^\infty \frac{(-1)^k 
(m+n+k+1)_k}{k! \,\,\Gamma(m+k+1)\,
\Gamma(n+k+1)}\, \left(\frac{z}{2}\right)^{m+n+2k} \label{pair}
\eeq

The generalized hypergeometric series  $_pF_q$ are defined by \cite{bateman}: 
\beq
_pF_q(a_1,\cdots,a_p;b_1,\cdots,b_q;z)=\sum_{k=0}^\infty 
\frac{(a_1)_k \cdots (a_p)_k}{(b_1)_k\cdots (b_q)_k}\cdot
\frac{z^k}{k!}\label{pq}
\eeq
where the Pochhammer  symbol  $(a)_k$ is defined by
\beq
(a)_0=1 \quad ,\quad 
(a)_k=\frac{\Gamma(a+k)}{\Gamma(a)} = a (a+1)\cdots (a+k-1)
\eeq
and $p \leq q+1$. Define a differential operator $\delta=z\frac{d}{dz}$.
The function (\ref{pq}) satisfies the differential equation
\beq
\left[\delta (\delta +b_1-1)\cdots (\delta +b_q -1) - z(\delta +a_1)
\cdots (\delta +a_p)\right] y(z) = 0
\eeq
For $p=q+1$ this equation is Fuchsian with three regular singular points
at $0$, $1$ and $\infty$. 
\\

\noindent{\Large \bf Appendix B: The WLW approach for arbitrary mass}
\\

\noindent Starting from an observation of C. Vohwinkel,
L\"uscher and Weisz have developed a powerful algorithmic 
method for the numerical calculation of the massless lattice Green
function in four dimensions \cite{luwe2}. Their method applies immediately
to any dimension. Here I show that this
method generalizes to arbitrary mass and anisotropies. 
A conserved quantity for the two-dimensional Green function at
the massless  and  decoupling points is also derived. 

An integration by parts of the left-hand side of 
(\ref{byp}) yields  the right-hand side: 
\bea
& & \alpha_j\, \left( G^{(d)}_{\pm}(\vec{x}+\hat{e}_j|\vec{\alpha},\be)
- G^{(d)}_{\pm}(\vec{x}-\hat{e}_j|\vec{\alpha},\be)\right) = - x_j
{\cal H}(\vec{x}|\vec{\alpha}) \; , \quad j=1,\cdots,d \label{byp}\\
& & {\cal H}(\vec{x}|\vec{\alpha}) = \int_{-\pi}^{\pi}\cdots \int_{-\pi}^{\pi}
\frac{d^d\vec{q}}{(2\pi)^d} \exp(i\vec{q}.\vec{x})\, \ln\left( 
2\be-2\sum_{j=1}^d \alpha_j\cos q_j \mp i\epsilon \right)
\eea
Equation (\ref{greq}) allows one to find another expression
for $\cal H$:
\beq
{\cal H}(\vec{x}|\vec{\alpha}) = \frac{2}{\sum_{j=1}^d x_j}\left(
\sum_{j=1}^d \alpha_j G^{(d)}_{\pm}(\vec{x}-\hat{e}_j|\vec{\alpha},\be)
- \be \, G^{(d)}_{\pm}(\vec{x}|\vec{\alpha},\be)\right)
\eeq
provided $\sum_{j=1}^d x_j\not= 0$. This gives the value of 
$G^{(d)}_{\pm}(\vec{x}+\hat{e}_j|\vec{\alpha},\be)$ in terms of  
$G^{(d)}_{\pm}(\vec{x}|\vec{\alpha},\be)$ and 
$G^{(d)}_{\pm}(\vec{x}-\hat{e}_k|\vec{\alpha},\be)$. 
The repeated use of these recurrence relations, coupled with 
the $\pm x_j$ invariance, shows that $G^{(d)}_{\pm}(\vec{x}|\vec{\alpha},\be)$
is a linear combination of the  $2^d$ values corresponding to  $x_j=0,1$. 
Note that all the vertices of  the unit hypercube 
are needed when the anisotropies are  arbitrary. 
These $2^d$ values can be calculated numerically,
and the particular ``$\pm$'' branch obtained 
depending on the given value of $\be$ in the complex plane.  
This generalizes the approach developed in \cite{luwe2}. 

One can  look for additional 
conserved quantities as was  done in \cite{luwe2}. However this method  
depends  rather strongly 
on the dimension. For the isotropic two-dimensional case, define 
\beq
g_0(n)=G^{(2)}_\pm((n,0)|\be) \quad , \quad \quad 
g_1(n)=G^{(2)}_\pm((n,1)|\be)\; , \quad\quad n\geq 0
\eeq
The Green  equation gives
\beq
g_0(n+1)+g_0(n-1)+2 g_1(n)-2\be\, g_0(n) =0 \; , \quad\quad n\geq 1 \label{eq1}
\eeq
and an equation inferred from the above approach is
\beq
g_1(n+1) = \frac{2n}{n+1}\left(\be\, g_1(n)-g_0(n)\right)-\frac{n-1}{n+1} 
g_1(n-1) \; , \quad\quad n\geq 1 \label{eq2}
\eeq
One can look for a conserved quantity in the following form
\bea
C(n) &=&  n g_0(n) + a_1 n g_1(n) + b_0 (n-1) g_0(n-1) + b_1 (n-1) g_1(n-1)\\
\phantom{C(n)} & & + c_0  g_0(n) + c_1  g_1(n)+ d_0  g_0(n-1) + d_1  
g_1(n-1)\; , \quad\quad n\geq 1 \nonumber
\eea
Using (\ref{eq1}) and (\ref{eq2}) one finds that $C(n)$ is independent 
of $n$ provided 
\beq
b_0=c_0=d_0= -1\, , \;  a_1=-b_1= \frac{1}{\be-1}\, , \; c_1=d_1=0
\eeq
and $\be=0$ or $\be=2$. This form does not allow for other values
of $\be$, but a conserved quantity at  $\be=-2$ can be obtained from 
the one for $\be=2$ through the parity symmetry. The new quantity
is {\it a priori} conserved for $n$ odd and $n$ even separately. 
It would interesting to find out
whether arbitrary values of $\be$ 
accommodate conserved quantities. 

Note that $\be=0,\pm 2$ are the three singular values. Therefore
$C(n)$ may not be well-defined. However, using the explicit
expression of section 4, and taking the limit $\be\rightarrow 0$, 
one finds that the infinities cancel exactly, leaving
$C_{\pm}\equiv C_{\pm}(n)= \pm\frac{i}{\pi}$ for all $n\geq 1$.
For the half-sum the conserved quantity is therefore $C=0$.
The situation at $\be=2$ is similar. 
The conserved quantities can be  
finite through cancellations, and the  divergence 
corresponds to a solution of the homogeneous equation
and can therefore be dropped. (See also 
the remark in the conclusion of \cite{luwe2} 
concerning this conserved quantity, and \cite{shin}.) 
\\

\noindent{\bf Note added:} Lattice Green functions also arise 
in the study of the statistical mechanics of the 
spherical model \cite{gsjoyce}. Complex temperature singularities of 
this system were studied in \cite{butera}. Lattice Green functions
were also examined  for the cases 
where factorizations in two complete elliptic integrals 
occur \cite{montaldi}. 
I would like to thank P. Butera for bringing to my attention 
the four works cited in this note.

\end{document}